# EFFICIENT AND RELIABLE HYBRID CLOUD ARCHITECTURE FOR BIG DATABASE

Narzu Tarannum and Nova Ahmed

Department of Computer Science and Engineering, North South University, Dhaka, Bangladesh

## ABSTRACT

*The objective of our paper is to propose a Cloud computing framework which is feasible and necessary for handling huge data. In our prototype system we considered national ID database structure of Bangladesh which is prepared by election commission of Bangladesh. Using this database we propose an interactive graphical user interface for Bangladeshi People Search (BDPS) that use a hybrid structure of cloud computing handled by apache Hadoop where database is implemented by HiveQL. The infrastructure divides into two parts: locally hosted cloud which is based on "Eucalyptus" and the remote cloud which is implemented on well-known Amazon Web Service (AWS). Some common problems of Bangladesh aspect which includes data traffic congestion, server time out and server down issue is also discussed.*

## 1. INTRODUCTION

Cloud computing is an emerging concept for Computer network arena. Data distribution, authentication and authorization [1] are major challenges to implement an application based on public database. In our work we are consider national ID database, prepared by election commission of Bangladesh [21]. Using these database we are propose an interactive web page for Bangladeshi People Search (BDPS) that will use the hybrid structure of cloud computing. This structure is divided into two parts. One infrastructure will be locally implemented using open source "Eucalyptus" [9], [16] and the other part of the infrastructure will be implemented on Amazon Web Service (AWS) [13] cloud. In a country like Bangladesh power failure and as well as internet connection failure are common problem. Data traffic congestion, SQL server time out issue and as a result server down is very frequent for any kind of national level searching issue like Higher Secondary school Certificate, Secondary School Certificate result publishing, TIN registration issue etc.

National level information access through database is an international challenge. This happened because of the risk of single server failure. To defend these problems we propose the Hybrid cloud Structure for BDPS which will be handled by apache Hadoop [12], [14], [20]. Our local elastic cloud would be sufficient for handling regular query and updates. But External AWS elastic cloud will ensure backup and the 7X24 service. Apache Hadoop [14] is an open-source software framework that supports the running of applications on large clusters of commodity hardware. Map-reduce [12] paradigm and distributed file systems are designed to handle node failures automatically in Hadoop Framework. It enables applications to work big data. To address the [1] authentication and authorization issue and make EC's Database in more effective, efficient and useful, we consider following ideas: 1. Everyone will have a password to access their information and take a printout in a specific format to use in official purpose. 2. Everyone can only check others information by entering info. 3. Academic, Job information, Criminal record can be entered and verified by same database. Figure 1 is showing the hybrid infrastructure for handling big data.

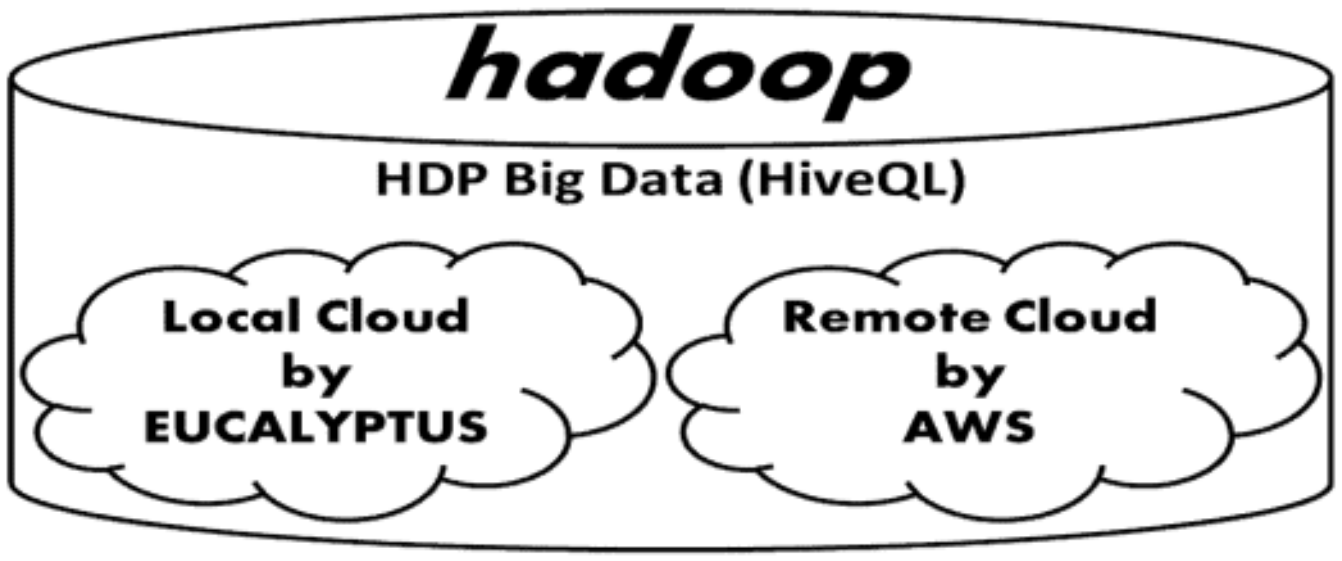

Figure 1. Hybrid Structure of Cloud for Handling Big Data

This paper is formatted in following way: - section II discusses about literature review, section III describes the motivation; we proposed architecture in section IV, section V describes about our implemented prototype and section VI discusses about evaluation.

## 2. LITERATURE REVIEW

We studied and reviewed some articles before designing and developing our system. Buyyaa et al. [1] have proposed architecture for market-oriented allocation. Distributed of resources within Clouds are also discussed in their literature. They discussed on meta-negotiation infrastructure for global Cloud exchanges which can provide high performance content delivery via Storage Clouds. Various Cloud efforts to reveal its emerging potential for the creation of third-party services is also discussed. The architecture, service model are based on this discussion.

Makhija et al. [2] discussed their methods of data security and privacy including different existing cloud components and methods. According to their review, they found many limitations on security mechanisms, lack in supporting dynamic data operations, data integrity. Considering all these limitations they proposed methods for ensuring data authentication using Third Party Auditor (TPA). Third Party Auditor is kind of inspector. TPA audits the data of client. Our proposed system would consider TPA for ensuring security and privacy issue. A message authentication codes (MACs) algorithm discussed by Mohta et al. [3] can be used to protect the data integrity and dynamic data operations. Provable data possession (PDP) method by Wang et al. [4] would be considered to ensure possession of data files on untrusted storages. Public key based homomorphic authenticator by Cong Wang et al. [5] and a schema “Proof of retrievability” for large files using “sentinels” proposed by Juels et al. [6] also considered for confirming the authentication and authorization issue in our system.

Ahmad et al. [7] presented a comprehensive analysis of cloud computing. They find the cloud concepts and demonstrate the cloud landscape vendors, growth of cloud computing, user concern about cloud security and worldwide web security revenue 2009 to 2015. They focused on the basic way of cloud computing development, growths and common security issues arising from the usage of cloud services. Their business model is considered to develop our business model.

Trancoso and Angeli [8] presented a brief description of GridArchSim, a computer architecture simulation environment that uses a database archive to reduce simulation latency and the Grid platform to increase the throughput of the simulations. This system is still under implementation. The system is going to be used for both research and education.

# 3. MOTIVATION

Cloud computing [10] is a universal term for anything that demands delivering hosted services over the Internet. A shared infrastructure of cloud must be hosted on the internet. Cloud computing platform works like a utility. This means need to pay only for required services or resources that would be used. Resource allocation can be adjusted. Day by day, the amount of data stored at companies like Google, Yahoo, Facebook, Amazon or Twitter has become incredibly huge. Handling big data in relational database became a new challenge. Thus, the necessity arises to make architectural change in web applications and databases in cloud because the advantage of the scalability provided by the cloud can’t be ignored.

Cloud computing concept appears as combination characteristic of clusters and grids, with its own attributes like storage support. Cluster is a kind of local area network which actually a single integrated computing used for centralized job management and system scheduling. Grid is a decentralized distributed job management, scheduling and computing network with super virtual computer. Cloud computing provides Software-as-a-Service (SaaS), Platform-as-a-Service (PaaS), Infrastructure-as-a-Service (IaaS), hardware virtualization, dynamically compo sable services with Web Service interfaces and utility which led to growth of cloud computing. Because services provided to the users without reference of infrastructure on which these are hosted. Cloud computing strongly support for creating 3rd party and can adopts the Service oriented architecture, cut costs, and provide quality service and support of intensive parallel computing.

Though cloud support in a remote location client server application service. We considered AWS and Windows Azure for Remote cloud implementation. Amazon Web Services (AWS) [13] is a collection of remote computing services that together make up a cloud computing platform, offered over the Internet by Amazon.com. AWS is located in 8 geographical regions. Amazon.com becomes popular when it introduced Elastic compute cloud.

Open Source has a number of advantages like leverage the work of a growing community, works though distributed hardware infrastructures, possible to deploy at service providers and on-premise, customized to fit individual needs etc. So we considered Eucalyptus and Openstack for designing our model as both of them are open source. Eucalyptus [16] allows an organization to build self-service, elastic clouds inside its data centre using existing IT infrastructure. Openstack [17] is a collection of open source components to deliver public and private IaaS clouds whose components are Nova, Swift, Glance, Keystone, and Quantum.

From the above discussion we found that AWS and Windows Azure are the options to implement Remote cloud. Unfortunately the service of Windows Azure is not available in Bangladesh and the services are not only expensive but also limited comparing to AWS for our application. AWS made a number of resources available to the researchers. We used some of them for our experiment which includes EC2, S3 etc. Different types of instances are verified for our application.

For local cloud, a comparative study by [25] Sonali Yadav is considered where the characteristics and performance is observed for Eucalyptus, Openstank and Opennebula. From that study we found that Eucalyptus would be better option for our proposed architecture because Eucalyptus provides an EC2 -compatible cloud Computing Platform and S3-compatible Cloud Storage thus its services are available through EC2/S3 compatible APIs. Eucalyptus can leverage (drag) a heterogeneous collection of virtualization technologies within a single cloud, to incorporate resources that have already been virtualized without modifying their configuration.

# 4. ARCHITECTURE OF HYBRID STRUCTURED CLOUD

We considered national ID database for searching Bangladeshi People in different purpose. An interactive web based application prototype by using hybrid structure of cloud computing has implemented in our research which is based on Hadoop with HortonWorks Data Platform (HDP). We used four elastic (EC2) nodes that are installed on Amazon Web Service (AWS). All the nodes including the head node is implemented on CentOS verstion-6.3 operating system. To address the authentication we also enabled public key and private key tool.

## 4.1. Structural Description

In our research we proposed a hybrid structure of cloud computing as depicted in figure 2. This structure is divided into two parts. One infrastructure will be locally implemented by using "Eucalyptus" and the other part of the infrastructure will be implemented in well-known Amazon Web Service (AWS) cloud. On top of this infrastructure Hadoop framework would be used to implement the system. In our structure the solid lined servers are representing the "always on" server. In local elastic cloud part those servers will be used for query handling requested by the users and in External AWS cloud those server will be used for backup and mirroring. In local elastic cloud part dashed line servers will be used as elastic computer which will be automatically "UP" as needs' basis. The number of server will depend on the number of query request. In External AWS cloud the dashed server will be used is case of overflow request and in case of local cloud infrastructure failure. Any kind of Linux server can be used for this implementation. We used CentOS in our prototype cloud and Hadoop implementation. [23], [24] HiveQL would be the preferable database for our proposed system but we used MySQL for our prototype. Different type of devices around Local Elastic Cloud indicates that our system would support entire computing platform.

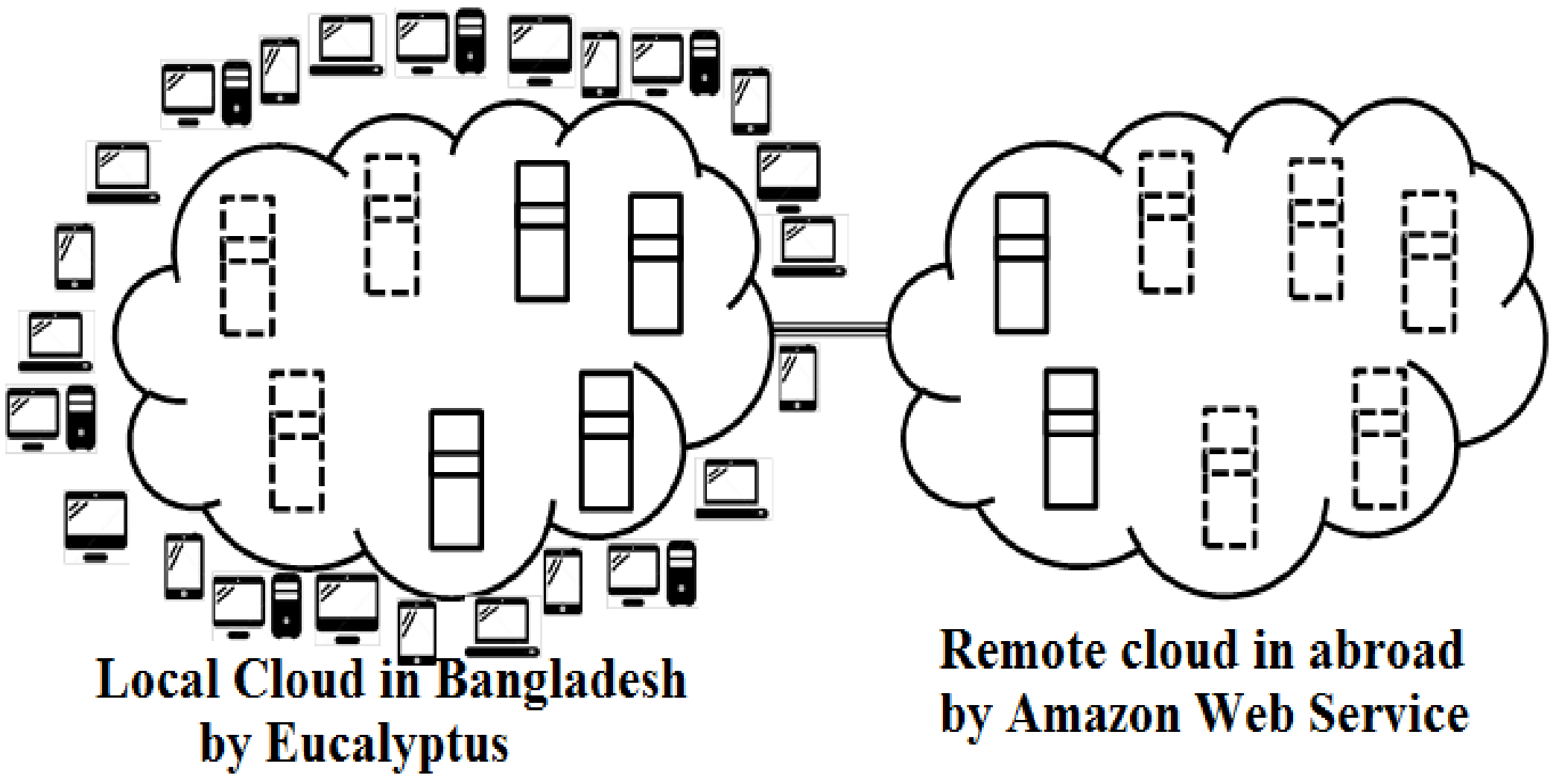

Figure 2. Block Diagram of BDPS System

Figure 3 is showing the architecture of our proposed system. It shows that the Database is stored in the hybrid cloud which is made by Local and Remote cloud. Hadoop framework contains both structure and the Database. Data input authority and User can access this database by a user interface which is connected to Hadoop with secured communication protocol. The selection criteria for local and remote cloud are also shown in our architecture.

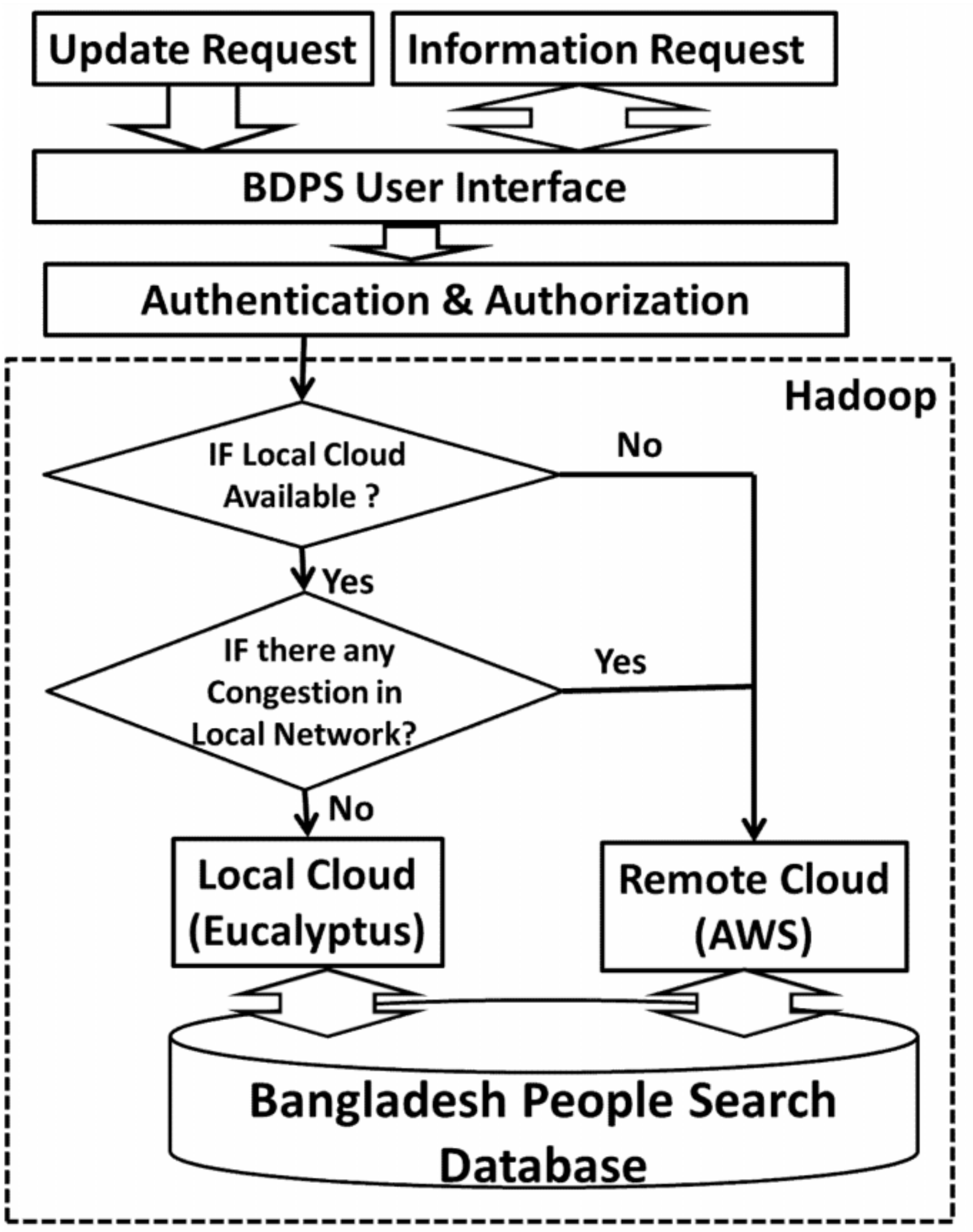

Figure 3. Architecture of BDPS System

## 4.2. Technology of implementation

An open-source software framework Apache Hadoop [14] supports data distributed applications. In this framework large and segmented hardware is used to run an application. It also supports commodity hardware infrastructure. Hadoop provides reliable data motion to all of its applications. This framework implements map-reduce [12] computational paradigm. In map-reduce paradigm, an application is divided into many small pieces of work and they are distributed to the nodes in the cluster for execution. Pig and Hive are Higher-level languages over Hadoop that generate MapReduce programs.In addition, Hadoop provides a distributed file system that stores data on the compute nodes, providing very high aggregate bandwidth across the cluster. Map-reduce and the distributed file systems are not only designed for running the application efficiently but also to handle node failures automatically. Typical hadoop cluster is shown in figure 4.

The challenges of cheap nodes fail, commodity network, programming distributed systems are solved by building fault-tolerance into system, push computation to the data and data-parallel programming model: users write “map” & “reduce” functions, system distributes work and handles faults. MapReduce programming model hides the complexity of work distribution and fault tolerance. The main principal of this design is to make the system scalable and cheap. The main features of Hadoop Distributed File System (HDFS) are single namespace for entire cluster and replicates data 3x for fault-tolerance.

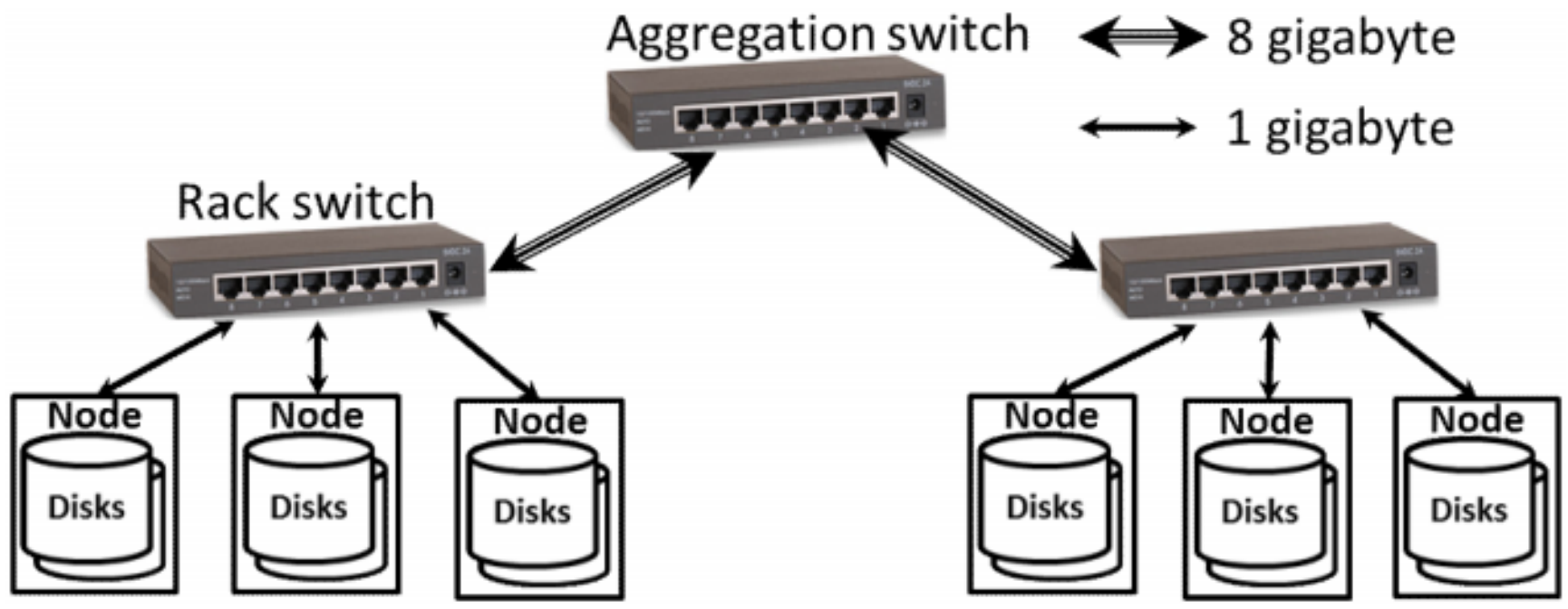

Figure 4. Typical hadoop cluster

The Hortonworks Data Platform [20] is a framework to implement Apache Hadoop. The platform includes various Apache Hadoop projects including the Hadoop Distributed File System, MapReduce, Pig, Hive, HBase and Zookeeper and many other additional components. The platform is designed to deal with big data from many sources and formats.

The National ID database of Bangladesh contains 31 information of voter which includes picture and figure print. The total number of voter is almost 95 million on November, 2013. National ID number of Bangladesh contains 13 digits. The structure is "DDRTTUUSSSSSS". "DD" is using for District code, "R" for R.M.O code, "TT" for Thana code, "UU" for Union code and the last "SSSSSS" six digits are a unique number for each citizen of the country. Figure 5 is showing a sample of national ID card.

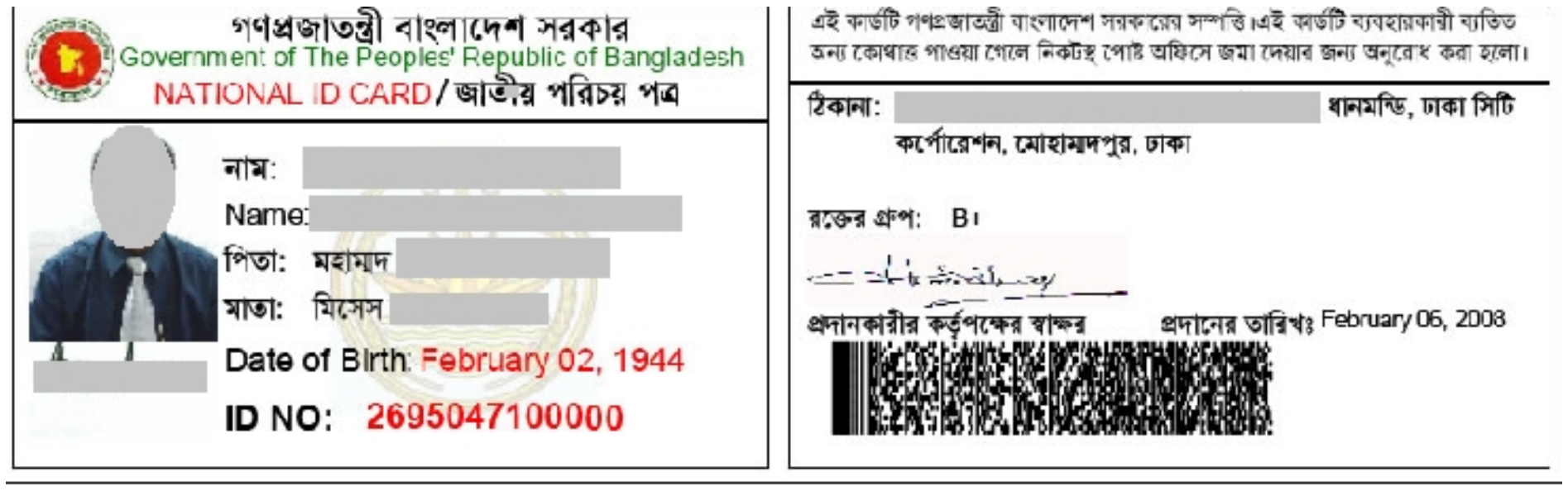

Figure 5. Sample of National ID card of Bangladesh

For handling this huge database, an interactive web based application is proposed using hybrid structure of cloud computing. A prototype of this infrastructure has implemented in our research which is based on [14] Hadoop with [20] HortonWorks Data Platform (HDP). We used four elastic (EC2) nodes that are installed on Amazon Web Service (AWS). [19] All the nodes including the head node is implemented on CentOS verstion-6.3 operating system. To address the authentication we also enabled public key and private key tool. This system contains entire tools of hadoop which is shown in Figure 6. We discuss some of components in next part of our discussion related of this frame work.

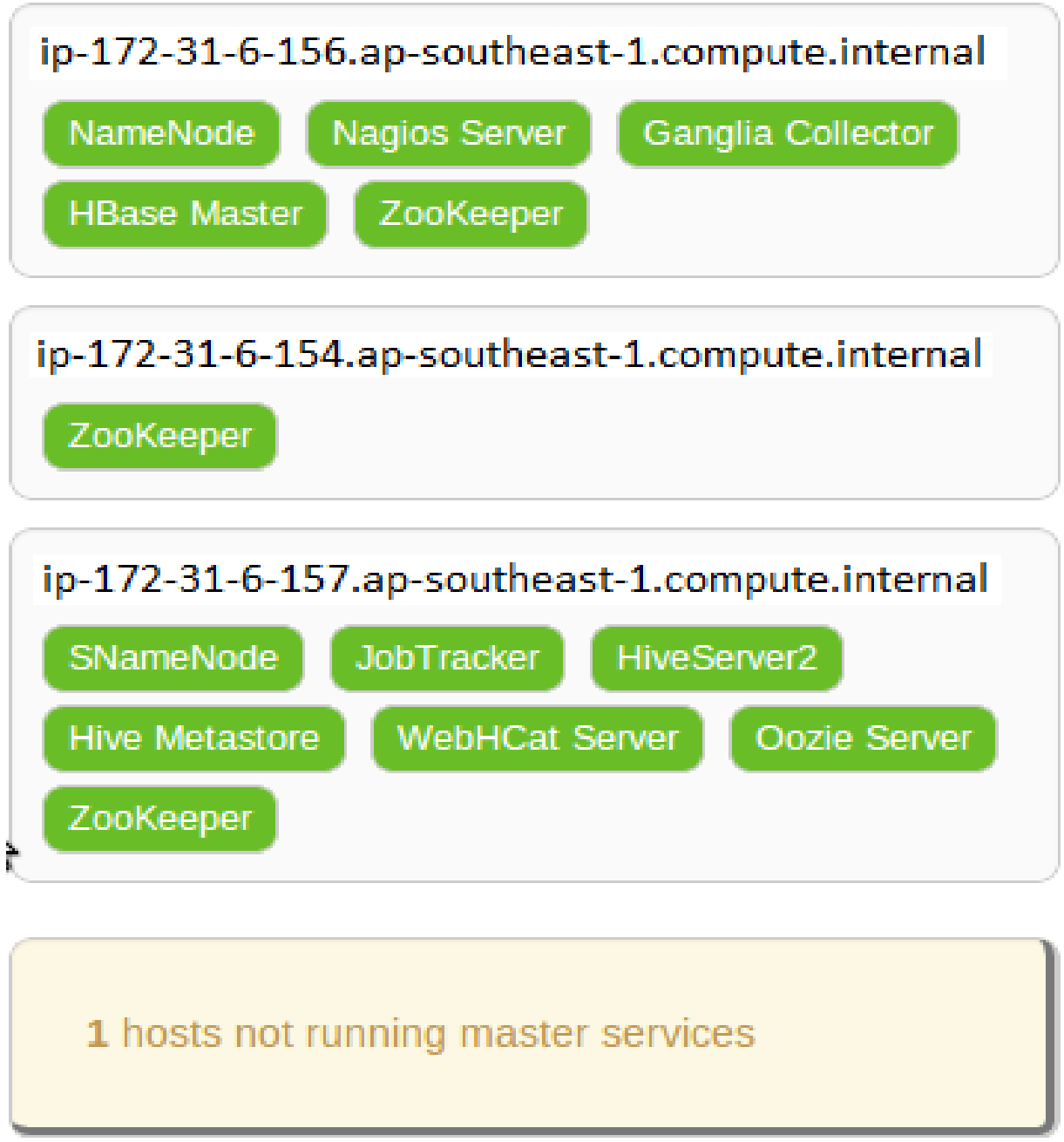

Figure 6. Four Nodes having Hadoop Components

# 5. SYSTEM PROTOTYPE IMPLEMENTATION

## 5.1. Deployment model for BDPS

BDPS will make the database available to the general people by using the public cloud. Here Election Commission or a National Data Center will be the service provider. These services are free for General people to access their information. Election Commission or a National Data Center can introduce a fee for new registration and update process. Since a number of agencies will use this database, a huge amount of revenue is also possible. BDPS will also offer a pay-per-use model for the corporate user who will use this database frequently for information verification.

BDPS own and operate the infrastructure and offer access only via Internet. 3G internetwork will be available in almost everywhere. So, this service would be visible in every mobile device and would be popular with in very short time.

## 5.2. Service Model for BDPS

Since PaaS helped to run the application on the web and also provide application development toolkits, we choose PaaS as a service model for BDPS. The user of BDPS does not manage or control the underlying cloud infrastructure including network, servers, operating systems, or storage, but has control over the deployed applications.

One important issue in our application was the visibility. Since it is general software and everyone have access to this database, visibility of other's information became an important issue. A unique solution is proposed and implemented to address this issue in our research as follow: 1. everyone will have a password to access their information and take a printout in a specific format to use in official purpose, 2. everyone can only verify other's information by entering known information and 3. Academic, Job information, Income tax, Criminal record can be entered and verified.

## 5.3. DATABASE

In our prototype, we used five tables for five kinds of data. Our current election commission [22] stores 31 information for each voter. We kept all those data in information table. Academic Information, Job Record, Bank Account information and Criminal Record stored in other four tables. We implemented our prototype database on MySQL. Tables I is showing the name of fields for corresponding tables which is based on the Voter ID application form of election commission Bangladesh shown in Figure 7. Partial structures of entire database tables are shown in Figure 8. Bold-italic field means primary key in the figure.

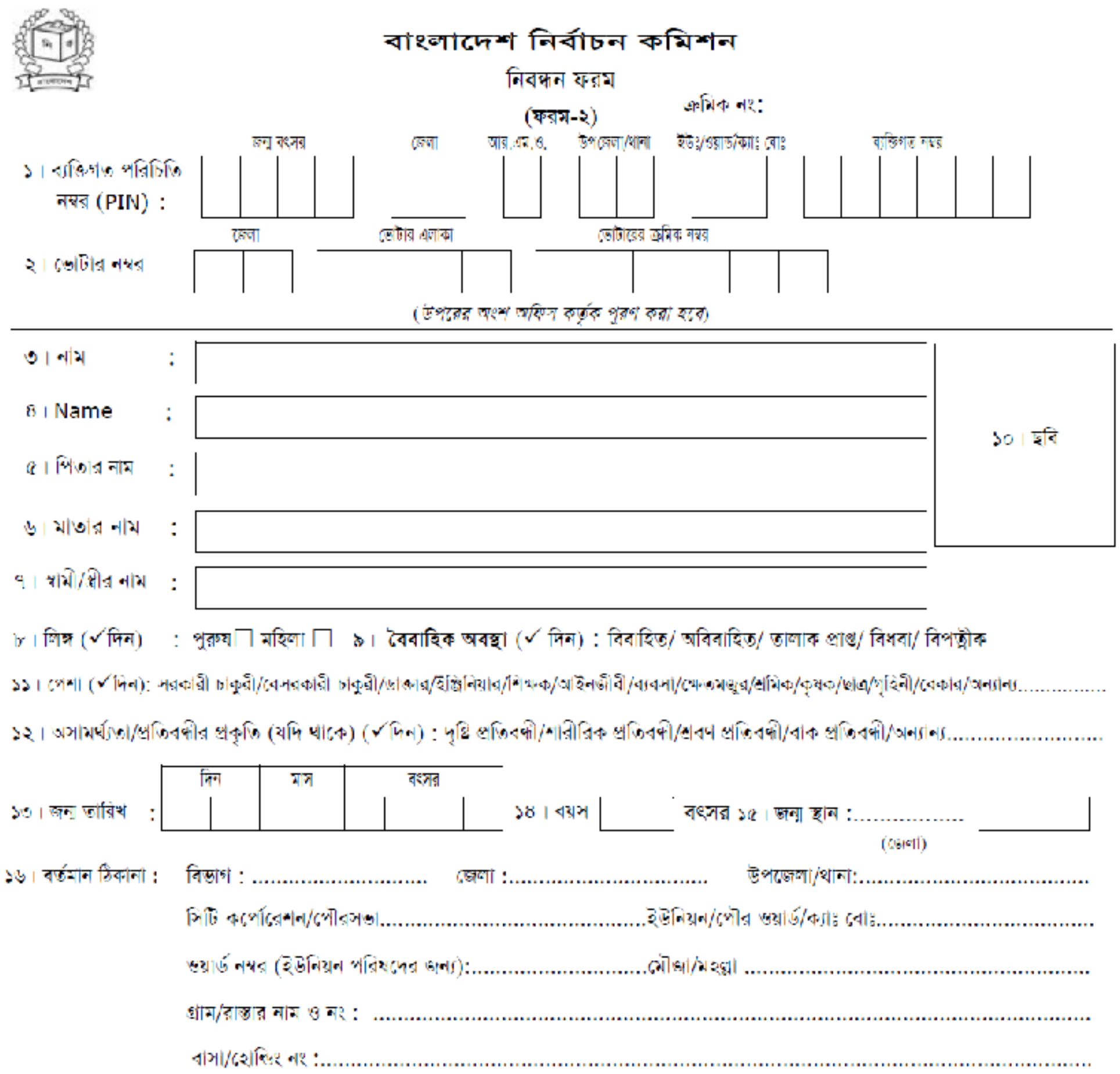

বাংলাদেশ নির্বাচন কমিশন

নিবন্ধন ফরম

(ফরম-২)

ক্রমিক নং:

১। ব্যক্তিগত পরিচিতি নম্বর (PIN) : জন্ম বৎসর / জেলা / আর.এম.ও. / উপজেলা/থানা / ইউঃ/ওয়ার্ড/ক্যাঃ বোঃ / ব্যক্তিগত নম্বর

২। ভোটার নম্বর : জেলা / ভোটার এলাকা / ভোটারের ক্রমিক নম্বর

(উপরের অংশ অফিস কর্তৃক পূরণ করা হবে)

৩। নাম :

৪। Name :

৫। পিতার নাম :

৬। মাতার নাম :

৭। স্বামী/স্ত্রীর নাম :

১০। ছবি

৮। লিঙ্গ (✓ দিন) : পুরুষ ☐ মহিলা ☐ ৯। বৈবাহিক অবস্থা (✓ দিন) : বিবাহিত/ অবিবাহিত/ তালাক প্রাপ্ত/ বিধবা/ বিপত্নীক

১১। পেশা (✓ দিন): সরকারী চাকুরী/বেসরকারী চাকুরী/ডাক্তার/ইঞ্জিনিয়ার/শিক্ষক/আইনজীবী/ব্যবসা/দিনমজুর/শ্রমিক/কৃষক/ছাত্র/গৃহিনী/বেকার/অন্যান্য...............

১২। অসামর্থ্যতা/প্রতিবন্ধীর প্রকৃতি (যদি থাকে) (✓ দিন) : দৃষ্টি প্রতিবন্ধী/শারীরিক প্রতিবন্ধী/শ্রবণ প্রতিবন্ধী/বাক প্রতিবন্ধী/অন্যান্য.........................

১৩। জন্ম তারিখ : দিন / মাস / বৎসর ১৪। বয়স বৎসর ১৫। জন্ম স্থান :................ (জেলা)

১৬। বর্তমান ঠিকানা : বিভাগ : ............................ জেলা :................................ উপজেলা/থানা:.....................................

সিটি কর্পোরেশন/পৌরসভা.........................................ইউনিয়ন/পৌর ওয়ার্ড/ক্যাঃ বোঃ...................................

ওয়ার্ড নম্বর (ইউনিয়ন পরিষদের জন্য):..........................মৌজা/মহল্লা .......................................................

গ্রাম/রাস্তার নাম ও নং : ..........................................................................................................

বাসা/হোল্ডিং নং :..........................................................................................................

Figure 7. Application form of Election Commission

| information | | | |
|---|---|---|---|
| **Column** | **Type** | **Column** | **Type** |
| ***DID*** | int(11) | Voter_ID | varchar(50) |
| PIN_ID | varchar(20) | Name | varchar(50) |

| job_record | |
|---|---|
| **Column** | **Type** |
| ***Job_ID*** | int(11) |
| National_ID | varchar(20) |
| Job_title | varchar(20) |

| education | |
|---|---|
| **Column** | **Type** |
| ***Education_ID*** | int(11) |
| National_ID | varchar(20) |
| Degree_name | varchar(20) |

| bank_acc_loan | |
|---|---|
| **Column** | **Type** |
| ***Bank_ID*** | int(11) |
| National_ID | varchar(20) |
| Account_name | varchar(50) |

| criminal_record | |
|---|---|
| **Column** | **Type** |
| ***Record_ID*** | int(11) |
| National_ID | varchar(20) |
| Record_no | varchar(20) |

Figure 8. GUIs for information verification

Table 1. FIELDS OF DATABASE.

| Table Name | Fields |
|---|---|
| Information | DID, PIN_ID, Voter_ID, Name, English_name, Father_name, Mother_name, Spouse_name, Gender, Merital_status, Picture, Qualification, Special, Date_of_birth, Birth_district, Present_address, Permanent_address, Voter_area,,Occupation, Specification_sign, B_group, TIN, License, Passport, IRIS_DNA, Phone, Nationality, F_print, Death_date |
| criminal_record | Record_ID, National_ID, Record_no, Case_no, Type, Place, Police_station, Date, Status, Details |
| bank_acc_loan | Bank_ID, National_ID, Account_name, Bank_name, Branch_name, Account_no, Card_no, Account_type, Date, Remarks |
| Education | Education_ID, National_ID, Degree_name, Year, Registration_no, Roll_no, Result, Marks, Remarks |
| job_record | Job_ID, National_ID, Job_title, Institute, Address, Designation, Joining_date, Departure _date, Remarks |

## 5.4. GUIs

Our GUIs are written in PHP. First GUI in Figure 9 is the interface of entering data. We have shown a partial record which we entered in the database. By the second GUI checked with some valid and invalid information and third GUI shows the response of information verification. By this concept we can ensure the visibility of private information from others.

## 5.5. SOME SAMPLE SQLs

### 5.5.1. Insertion Query

We used PHP for entire SQL operations. SQL for Insertion information in the table which PHP code is similar for all other tables:

```
$sql="INSERT INTO information (DID, NID, E_name, Name, F_name,
M_name, . . . , Religion)
VALUES
(NULL, '$_POST[nid]', '$_POST[ename]', '$_POST[name]', '$_POST[fname]',
'$_POST[mname]' , . . . , '$_POST[religion]')";
```

### 5.5.2. Partial SQL for data verification

```
$result = mysqli_query($con, "SELECT * FROM information WHERE
NID='$_POST[nid]'");
while ($row = mysqli_fetch_array($result))
{ echo "For the National ID:"; echo $row['NID']; echo "<br>"; echo "<br>";
echo "<br>"; echo "Name:  "; echo $_REQUEST["ename"];
   if ($_REQUEST["ename"] == $row['E_name']) {
      echo ".......OK";
   }else { echo "XXXXXXXXXX...Wrong"; }
echo "<br>"; echo "Name in Bangla:  "; echo $_REQUEST["name"];
   if ($_REQUEST["name"] == $row['Name']) {
      echo ".......OK";
   } else {echo "XXXXXXXXXX...Wrong"; }
echo "<br>"; echo "Name in Bangla:  "; echo $_REQUEST["name"];
   if ($_REQUEST["f_name"] == $row['F_name']) {
      echo ".......OK";
   } else {echo "XXXXXXXXXX...Wrong"; }
echo "<br>"; echo "Name in Bangla:  "; echo $_REQUEST["name"];
   if ($_REQUEST["m_name"] == $row['M_name']) {
      echo ".......OK";
   } else {echo "XXXXXXXXXX...Wrong"; }
……
……
echo "<br>";echo "Religion:  "; echo $_REQUEST["religion"];
   if ($_REQUEST["religion"] == $row['Religion']) {
       echo ".......OK";
   } else { echo "XXXXXXXXXX...Wrong";}
  echo "<br>"; }
```

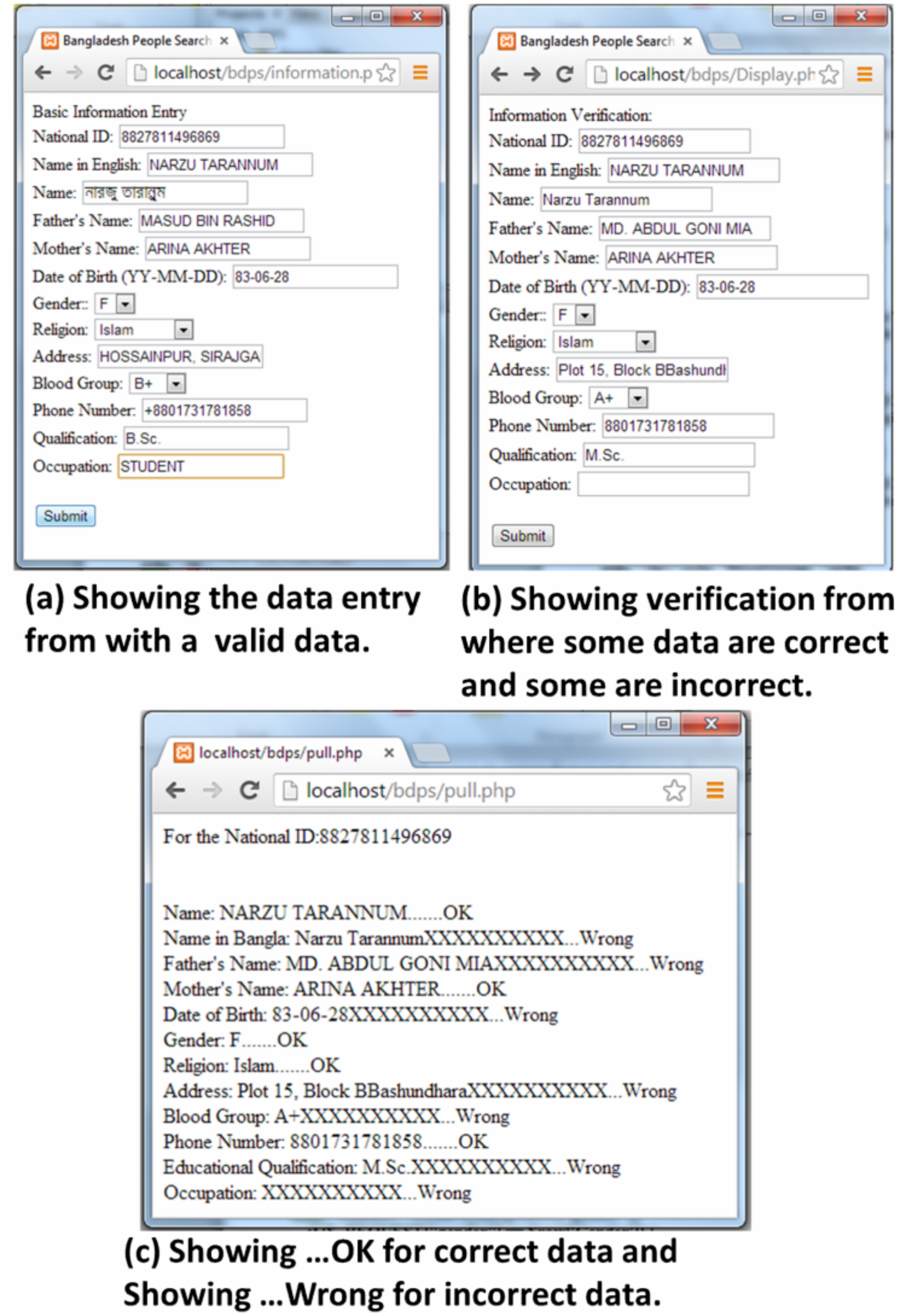

Figure 9. GUIs for information verification

## 6. EVALUATION

According to our proposal and considering the socio-economical aspect, we are expecting 100 thousand request per day in first year, 0.5 million request per day in 3nd year which will touch 1 million mark after five years' service. We projected these numbers by mainly considering the literacy rate of Bangladesh. We also considered that in first few years' educational information,

Job information and police record would be incomplete. Since the volume will be increased after every year, we would send the record in different backup file after death. So after 5 years the database volume will come to a stable position.

According to our projection, our local systems would be deployed on a 20-node cluster [11][12]. Each node would have a single 2.40 GHz Intel Core 2 Duo processor. 64-bit CentOS Linux (kernel version 6.3) would be used as Operating System. 4GB RAM with two 250GB SATA-I hard disks would be used as memory and storage. The nodes would be connected with Cisco Catalyst 3750E-48TD switches. The switches are linked together via Cisco StackWise Plus, which creates a 64Gbps ring between the switches.

To justify our proposal we have implemented and experimented with many tools of cloud computing. At the beginning we had simulated a cloud on [18] Cloudsim then we deployed Openstack and experiment some applications for performance study. After that we registered and experimented a number of applications in AWS which includes many kind of EC2 instances, S3 buckets etc. We lunched different type of instances like Windows, CentOS, Ubuntu etc. in different types of machine configurations. Using AWS service we deployed Hadoop on four node computing environment on Apache Ambari Server. Hortonworks big database was studied on this infer-structure. And lastly we implemented our prototype windows apache server and verified the data entry, visibility and authentication.

# 7. DISCUSSION

Large scale data handling is a major challenge in Bangladesh. Election commission (EC) of Bangladesh has a national database with 31 types of information for 95 million of voters. Huge cost is involved to maintaining this database which is supposed to use only at the time of election. Considering this point we proposed an application which would be useful to general people, government and non-government organizations along with EC. It will not only ensure the validity of data but also ensure the transparency. We proposed hybrid architecture with map reduce feature for a big database handling application. According to our proposal we can deploy the system on our own local cloud and can synchronize our data with remote cloud. This is an efficient and fall tolerant architecture. Service we proposed is “pay per use” which would support financially to maintain the database. The data and database is controllable and expandable according to the system’s requirement.

# ACKNOWLEDGEMENTS

We are acknowledging Amazon Web Service for their kind education grant for this research along with Election Commission of Bangladesh for their support.

## Authors

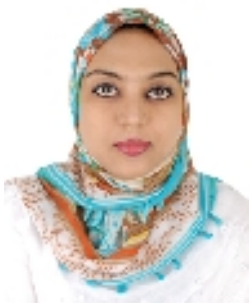

Narzu Tarannum is a M.Sc. student of the Department of Computer Science and Engineering of North South University. She completed her B.Sc. from South East University. Her Research interest includes Natural Language Processing, Networking and Communication along with cloud computing.

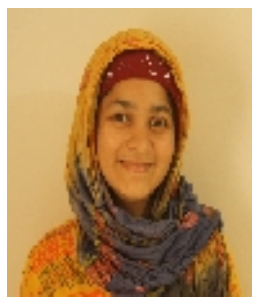

Dr. Nova Ahmed is the Assistant Professor the Department of Computer Science and Engineering of North South University. She completed her Ph.D. in MS in Computer Science from Georgia Institute of Technology. She finished MSc and B.Sc. in Computer Science from University of Dhaka. Her research interest includes Healthcare systems using Sensor Support, Unreliable Sensor Systems, Parallel and Distributed algorithms, Mobile Computing, High Performance Computing.